\documentclass[preprint,12pt]{elsarticle}

\usepackage{amssymb}

\usepackage{epsfig}
\usepackage{bm}
\usepackage{wrapfig}
\usepackage{ulem}
\usepackage{multirow}
\usepackage{color}

\newcommand{\Asla}{\ooalign{\hfil/\hfil\crcr{$A$}}}

\newcommand{\el}{{\cal L}}

\newcommand{\cT}{{\cal T}}

\newcommand{\psibar}{\mbox{$\overline{\psi}$}}

\newcommand{\tldJ}{\mbox{${\tilde J}$}}

\newcommand{\phibar}{\mbox{$\overline{\phi}$}}
\newcommand{\Vsla}{\ooalign{\hfil/\hfil\crcr{$V$}}}
\newcommand{\cM}{{\cal M}}

\newcommand{\vp}{\mbox{$\bm{p}$}}

\newcommand{\vbr}{\mbox{$\bm{r}$}}

\newcommand{\vA}{\mbox{$\bm{A}$}}
\newcommand{\vB}{\mbox{$\bm{B}$}}

\newcommand{\vsigma}{\mbox{$\bm{\sigma}$}}
\newcommand{\vepsi}{\mbox{\boldmath $\epsilon$}}

\newcommand{\valpha}{\mbox{\boldmath $\alpha$}}

\newcommand{\zr}{\mbox{${\bm r}$}}

\journal{Physics Letters B}

\begin{document}

\begin{frontmatter}



\title{Generation of photon vortex by synchrotron radiation from electrons in Landau states under astrophysical magnetic fields}


\author[Nihon,NAOJ]{Tomoyuki~Maruyama}

\address[Nihon]{College of Bioresource Sciences, Nihon University, Fujisawa 
252-0880, Japan}
\address[NAOJ]{National Astronomical Observatory of Japan, 2-21-1 Osawa, Mitaka, Tokyo 181-8588, Japan}

\author[QST,ILE]{Takehito~Hayakawa}
\address[QST]{National Institute for Quantum and Radiological Science and Technology, Tokai, Ibaraki 319,-1106, Japan}
\address[ILE]{Institute of Laser Engineering, Osaka University, Suita, Osaka 565-0871, Japan}

\author[Beihang,NAOJ,Tokyo]{Toshitaka~Kajino}
\address[Beihang]{Beihang University, School of Physics,\\
Int. Center for Big-Bang Cosmology and Element Genesis, Beijing 100083,
China} 
\address[Tokyo]{The University of Tokyo, Bunkyo-ku, Tokyo 113-0033, Japan}

\author[Soognsil]{Myung-Ki Cheoun}
\address[Soognsil]{Department of Physics, Soongsil University, Seoul 156-743, Korea}

\cortext[cor2]{maruyama.tomoyuki@nihon-u.ac.jp}
\cortext[cor1]{hayakawa.takehito@qst.go.jp}

\begin{abstract}
We explore photon vortex generation in synchrotron radiations from a spiral moving electron under a uniform magnetic field along $z$-axis using Landau quantization.
The obtained wave-function of the photon vortecies is the eigen-state of the $z$-component of the total angular momentum (zTAM).
In $m$-th harmonic radiations, individual photons are the eigen-state of zTAM of $m$${\hbar}$.
This is consistent with previous studies.
Using the presently obtained wave-functions we calculate the decay widths and the energy spectra under extremely strong magnetic fields of 10$^{12}$$\--$10$^{13}$~G, which are observed in astrophysical objects such as magnetized neutron stars and jets and accretion disks around black holes.
The result suggests that photon vortices are predominantly generated in such objects. 
Although they have no coherency it is expected that photon vortices from the universe are measured using a detector based upon a quantum effect in future. 
This effect also affects to stellar nucleosynthesis in strong magnetic fields.

\end{abstract}


\end{frontmatter}


\section{Introduction}

In the universe strong magnetic fields play important roles in various phenomena.
Strong magnetic fields in rotating massive stars contribute to magnetohydrodynamic (MHD) driven core-collapse supernova explosions and formation of magnetized neutron stars or black holes associated with an accretion disk and jets (collapsar) \cite{MacFadyen99, Ardeljan00, Takiwaki11}.
Neutron stars may be associated with strong magnetic fields of 10$^{12}$$\--$10$^{15}$~G and, in particular, extremely strongly magnetized neutron stars, so-called magnetars, are considered to be the sources for soft $\gamma$ repeaters and anomalous X-ray pulsars \cite{Mereghetti08} and to be the central engines of short Gamma-Ray Bursts (GRBs) \cite{Zhang10, McKinney13, Lei18}.
Collapsars with a magnetized accretion disk and jets are also a candidate for the origin of long GRBs.
Recent progress in $\gamma$/X ray astronomy enabled to measure linear (circular) polarization of gamma-rays from these astrophysical objects.
A high linear polarization of 80\% $\pm$ 20\% measured by the RHESSI satellite \cite{Coburn03} was reported, although it decreases to approximately 41 $^{+59}_{-44}$\% by re-analysis \cite{Wigger04}.
The linear polarization as high as 98\% $\pm$ 33\% in the prompt emission of GRB 041219A was measured by the SPI telescope onboard the INTEGRAL satellite \citep{Kalemci07}.
The GRB polarimeter onboard the IKAROS solar power sail measured the polarization degrees of 70\% $\pm$ 22\% for GRB 110301A and 
the polarization of 84 $^{+16}_{-28}$\% for GRB 110721A \cite{Yonetoku12}.\textbf{\emph{}}
As the generation mechanism for these high linear polarized $\gamma$ rays, some scenarios have been proposed; they are synchrotron radiations from relativistic electrons under strong magnetic fields \cite{Panaitescu00} and inverse Compton scattering on low energy photons with relativistic electrons \cite{Chang14}.
The origin of these highly polarized photons has been unresolved but synchrotron radiation from electrons in strong magnetic fields is one the candidates for the mechanism \cite{Mao18}.

The synchrotron radiation in quantum theory was first derived by a pioneering work \cite{Sokolov45,Sokolov53} and the quantum synchrotron radiation under various conditions were studied  \cite{Klepikov54,Bezchastnov88,Pavlov91,Bagrov,Dorofeyev,Bisnovatyi-Kogan,Wistisen15,Kruining19}.
The electron orbitals under magnetic fields are in Landau levels.
An electron in a Landau level transits to a lower lying Landau state through an emission of a photon, where the wave-function of the radiated photon depends on the initial and final electron wave-functions.
As a magnetic field strength becomes weaker, an average level spacing decreases so that the Landau levels become quasi-continues and classical calculation becomes a good approximation. 
However, in strong magnetic fields as high as 10$^{12}$$\--$10$^{15}$~G each level is separated in the energy space and thus calculated results based upon Landau quantization are largely different from those without that \cite{Bagrov,Dorofeyev,Bisnovatyi-Kogan}.

Generation and observation of light vortices \cite{Allen92} in the universe have been discussed \cite{Harwit03,Elias08,Berkhout08,TTMA11}.
It is suggested that light vortices are created around rotating black holes \cite{TTMA11}.
Light vortices carry large angular momenta \cite{Allen92} and the interactions on materials are different from that with standard (plane-wave) photons \cite{Jentschura11a, Harris15, Petrillo16, Sherwin17, Afanasev17, Taira17, Peshkov18, Maruyama19a, Lee19, Sherwin20}.
Light vortices are generated from synchrotron radiations \cite{Sasaki08, Bahrdt13, Hemsing13, Katoh17, Bogdanov18, Bogdanov19}.

Allen et al.~\cite{Allen92} pointed out that a single photon could have a vortex wave-function such as Laguerre-Gaussian in quantum level.
Katoh et al.~\cite{Katoh17} suggested that an $l$-th harmonic photon in synchrotron radiation from spiral moving electrons under uniform magnetic fields is a photon vortex carrying $l$$\hbar$ total angular momentum and that such photon vortices are naturally generated in astrophysical objects with strong magnetic fields.
However, the wave-function of photon vortices radiated from an electron under a uniform magnetic field has not been calculated by taking Landau quantization into account.
In extremely strong magnetic fields as high as 10$^{12}{\--}$10$^{15}$~G, which are observed in astrophysical objects, the difference between calculated results of synchrotron radiations with/without Landau quantization becomes large \cite{Bagrov,Dorofeyev,Bisnovatyi-Kogan,Maruyama16}.
The purpose of this paper is to present the wave-functions of photons radiated from electrons under strong magnetic fields 
calculated using Landau quantization and the fraction of photon vortices in synchrotron radiations in extremely strong magnetic fields
of up to 10$^{13}$~G.
We also discuss the possibility of detection of photon vortices from the universe
and its effects to stellar nucleosynthesis.

\section{Calculation and Result}

\begin{figure}[htb]
\begin{center}
{\includegraphics[scale=0.52]{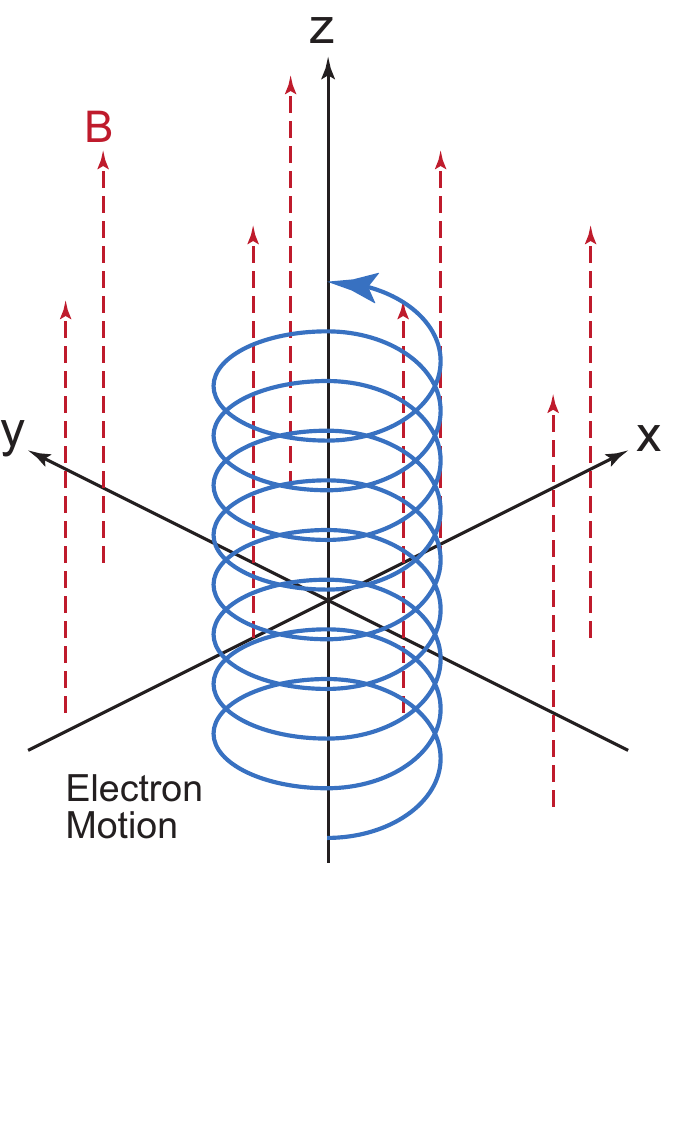}}
\caption{\small
Coordinate in the present study. We assume the uniform dipole magnetic field along the $z$-axis. The electron is in the spiral moving along the $z$-axis.}
\label{Coordinate}
\end{center}
\end{figure}

%
In the present study, we consider a uniform dipole magnetic field along $z$-direction, $\vB = (0, 0, B)$, and that an electron trajectory draws a circle in a plane perpendicular to the z-direction under this magnetic field (see Fig.~\ref{Coordinate}).
We use the natural unit $\hbar = c =1$.
The electron wave-function $\psi (\zr) $ in this system is obtained from the following Dirac equation:
\begin{eqnarray}
&& \left\{  {\valpha} \cdot (-i {\bm \nabla}_r + e{\mbox{$\bm{A}$}}) + \beta m_e - E \right\}  \psi (\zr) = 0 ,
\label{DrcEq}
\end{eqnarray}
where $\valpha$ and $\beta$ are the Dirac matrices, $\vA$ is an electro-magnetic vector potential, $E$ is the electron energy, $e$ is the elementary charge, and $m_e$ is the electron mass.
We choose the symmetry gauge with the vector potential being
$\vA = (-y, x, 0 )B/2$.
We make the scale transformation for $x$ and $y$ coordinates as 
$X (Y) = (\sqrt{eB/2}) x (y) $ ,
and define the operators as
\begin{eqnarray}
a &=& \frac{1}{2} \left( X - i Y + \nabla_X  - i \nabla_Y \right),
\nonumber \\
b &=& \frac{1}{2} \left( X + i Y + \nabla_X + i \nabla_Y \right)
\nonumber \\
a^\dagger &=& \frac{1}{2} \left( X + i Y - \nabla_X - i \nabla_Y \right),
\nonumber \\
b^\dagger &=& \frac{1}{2} \left( X - i Y - \nabla_X + i \nabla_Y \right) .
\end{eqnarray}
As well known, Hamiltonian of the two-dimensional harmonic oscillator and the operator of the $z$-component of orbital angular momentum $L$ are written as
\begin{eqnarray}
{\hat h}_{HO} &=& \frac{1}{2} \left( -\nabla_X^2 -\nabla_Y^2 + X^2 + Y^2 \right)
= a^\dagger a  + b^\dagger b + 1 ,
\nonumber
\\
{\hat L}_z &=&  - i X \nabla_Y + i Y \nabla_X =  a^\dagger a  - b^\dagger b .
\label{Op}
\end{eqnarray}
We write following equations in the cylindrical coordinate of  $\zr = (\zr_T, z)  = ( r_T \cos \phi, r_T \sin \phi, z)$.
A function $G_n^L$ is the eigen-state of the operators of (\ref{Op}) as
\begin{equation}
{\hat h}_{HO} G_n^L (\zr_T) = (2n + |L| + 1) G_n^L,
\quad
 {\hat L_z}G_n^L (\zr_T) = L G_n^L,
\end{equation}
\begin{equation}
G_n^L (\zr_T) 
= \sqrt{\frac{n!}{\pi (n + |L|)!} } e^{i  L \phi} r_T^{|L|} e^{ -r^2/2} \el_n^{|L|} ( r_T^2) e^{i L \phi} , 
\end{equation}
where $\el_n^{|L|}$ is the associated Laguerre function, $L$ is the z-component of the orbital angular momentum,  $p_z$ is the $z$-component of the momentum, and $n$ is the number of nodes.
$L$ satisfy a relationship of $L$ = $J$ + 1/2 where $J$ is the $z$-component of the total angular momentum (zTAM).
A solution of Eq.~(\ref{DrcEq}) in this system is known as \cite{Bagrov}
\begin{equation}
\psi (\vbr) = {\cal N} \left[ \frac{ 1 + \Sigma_Z}{2} 
G^{L-1}_{n^\prime} \left( \sqrt{ \frac{eB}{2}} \vbr_T \right)
+ \frac{ 1 - \Sigma_Z}{2} G^L_{n} \left( \sqrt{\frac{eB}{2} } \vbr_T \right) \right] 
U e^{ip_z z},
\label{Bord}
\end{equation}
where $n^\prime = n$ when $L\ge 0$ and $n^\prime = n-1$ when $L \le -1$, ${\cal N}$ is the normalization factor and $U$ is a  4-dimensional vector.
The wave-function is the eigen-state of zTAM.
The electron energy in a Landau level is given by $E = \sqrt{2 e B N_L + p_z^2 + m_e^2}$
where $N_L$ indicates the Landau level number defined as $N_L = (L + |L|)/2 + n$.

To obtain the Dirac spinor $U$ we rewrite Eq. (\ref{DrcEq}) with Eq.~(\ref{Bord}) as
\begin{eqnarray}
&& e^{-i p_z z}
\left\{  {\valpha} \cdot (-i {\bm \nabla}_r + e{\mbox{$\bm{A}$}}) + \beta m_e - E \right\}  \psi (\zr)  = 0
\nonumber \\
&=& 
\left[ \begin{array}{cccc} m_e - E & 0 & p_z & - i \sqrt{2 eB} a
\\ 0 & m_e - E  & i \sqrt{2 eB} a^\dagger & - p_z
\\ p_z & - i \sqrt{2 eB} a & - m_e -E & 0 
\\ i \sqrt{2 eB} a^\dagger & - p_z & 0 & -m_e - E  
\end{array} \right]
\left[ \begin{array}{cccc} G^{L-1}_{n^\prime}  & 0 & 0 & 0
\\ 0 & G^{L}_{n} & 0 \\  0 & 0 & G^{L-1}_{n^\prime}  & 0 
\\ 0 & 0 &  0 & G^{L}_{n} 
\end{array} \right] U 
\nonumber \\
&=& 
\left[ \begin{array}{cccc} G^{L-1}_{n^\prime}  & 0 & 0 & 0
\\ 0 & G^{L}_{n} & 0 \\  0 & 0 & G^{L-1}_{n^\prime}  & 0 
\\ 0 & 0 &  0 & G^{L}_{n}  \end{array} \right]  
\nonumber \\ && \qquad\qquad \times
\left[ \begin{array}{cccc} m_e - E & 0 & p_z & - i \sqrt{2 eB N_L} 
\\ 0 & m_e - E  &  i \sqrt{2 eB N_L}  & - p_z
\\ p_z & - i \sqrt{2 eB N_L}  & - m_e -E & 0 
\\ i \sqrt{2 eB N_L} & - p_z & 0 & -m_e - E  
\end{array} \right] U  .
\end{eqnarray}
By solving the above characteristic equation,
we can obtain the electron energy as 
\begin{equation}
U = \sqrt{\frac{E+m_e}{2E}}\left( \begin{array}{c} \chi_h \\  
\frac{ \tilde{\vp} \cdot \vsigma}{E+m_e}  \end{array} \right) ,
\end{equation}
where $E$ is the Landau energy. 
We finally obtain the wave-function as 
\begin{eqnarray}
\psi (\zr) &=&
\left\{ \frac{1 + \Sigma_z}{2} G_{n^\prime}^{L-1} \left( \sqrt{\frac{eB}{2}} \zr_T \right)
    + \frac{1 - \Sigma_z}{2} G_n^L \left( \sqrt{\frac{eB}{2}} \zr_T \right) \right\} 
\nonumber \\ && \qquad\qquad\qquad\qquad  \times
\sqrt{ \frac{E + m_e}{2 E} } \left[ 
\begin{array}{c}
\chi_h
\\
\frac{ \tilde{\vp} \cdot \vsigma}{E+m_e} \chi_h,
\nonumber
\end{array}
\right] \frac{ e^{i p_z z}}{\sqrt{R_z}}   ,
\label{DrWf}
\end{eqnarray}
\begin{equation}
\tilde{\vp} = ( 0, \sqrt{2 e B N_L}, p_z ) ,
\label{electron_function}
\end{equation}
where $R_z$ is the size of the system along the $z$-direction, 
$\vsigma \equiv (\sigma_x, \sigma_y, \sigma_z)$ is the Pauli matrix, $\chi_h$ is the two-dimensional Pauli spinor
satisfying $\sigma_z \chi_h = h \chi_h$ $(h=\pm 1)$,
$\Sigma_z = {\rm diag} (1,-1, 1, -1 )$.
Note that $n^\prime = n$ when $L \ge 0$ and  $n^\prime = n-1$ when $L \le -1$.
The form of $\chi_h$ is arbitrary when $N_L \ge 1$ because there are two degenerate states at fixed $L$ and $n$, whereas the state with $N_L=0$, which is so called the lowest Landau state, is not degenerate and its spinor is taken to be only
$^t \chi_{-1} = (0,1)$.
The wave-function for $n$ = 0 corresponds to the helical motion along the $z$-axis when $L$ $\ge$ 0, whereas the wave-function for $n$ $\ge$ 1 indicates the helical motion along an axis that is different from the initial axis \cite{RKubo65}.
Thus, we consider various node numbers for the final state and take only $n$ = 0 for the initial state.

Using the Coulofcolormb gauge $\bm{\nabla} \cdot \vA = 0$ with the photon as $A_0=0$, we obtain the wave-function propagating along the $z$-direction for the emitted photon as a solution of the Klein-Gordon equation in the cylindrical coordinate as stated previously.
We obtain $\vA$ as 
\begin{equation}
\vA_{ms} (\zr, t) = \vepsi_s J_{m-s} (q_T r) e^{i (m-s) \phi}  e^{i ( q_z z - e_q t) } ,
\end{equation}
\begin{equation}
\tldJ_M (\zr_T)  = J_M( q_T r_T) e^{i M \phi} ,
\label{bessel}
\end{equation}
where $m$ is the zTAM, $s$ is the helicity, $q_z$ is the $z$-component of the photon momentum, $e_q$ is the photon energy, $q_T = \sqrt{e_q^2 - q_z^2}$, $\vepsi_s = ( 1, is, 0 ) /\sqrt{2}$, and $J_{M}$ is the Bessel function.
This $\vA$ does not generally satisfy the gauge condition $\bm{\nabla} \cdot \vA = 0$
except for $e_q^2 - |q_z|^2 \ll 1$ in the para-axial limit.  
Then, we add a $z$-component of $A_z$ and rewrite $\vA$ as
\begin{equation}
\vA_{ms}  \propto 
e^{i (q_z z - e_q t) } 
\left[ - i q_z \tldJ_{m - s} (q_T r) ,  s q_z \tldJ_{m - s},  s q_T \tldJ_m \right]  .
\label{BessL0}
\end{equation}
which is consistent to the wave function given in Ref.~\cite{Bogdanov18} 
when $q_T \ll |q_z|$.
Because $\vA(s = +1)$ and $\vA(s = -1)$ do not satisfy the orthogonal relation,
we take the eigen-state wave-functions of the transverse magnetic (TM) state with  $\vA(s = +1)-\vA(s = -1)$ 
and those of the transverse electric (TE) state with  $\vA(s = +1)+\vA(s =-1)$ at fixed $m$ \cite{Jaurgui05}.
Finally, we obtain the wave-function of the radiated photons as
\begin{eqnarray}
\vA_{m}^{(TM)} &=&  \frac{1}{2 e_q} e^{i (q_z z - e_q t) } \left[ i q_z \left( \tldJ_{m+1} - \tldJ_{m-1}  \right) , 
q_z \left( \tldJ_{m+1} + \tldJ_{m-1} \right),  2 q_T \tldJ_{m}  \right] ,
\nonumber
\\
\vA_{m}^{(TE)}  &=&
 \frac{1}{2} e^{i (q_z z - e_q t) }  \left[ i \left( \tldJ_{m+1}  + \tldJ_{m-1}  \right) , 
 \left( \tldJ_{m+1} - \tldJ_{m-1} \right), 0  \right] .
\label{photon_function}
\end{eqnarray}
This photon wave-function is the eigen-state of zTAM,
where the zTAM of an emitted photon $m$ satisfies a relationship of $m$ = $J_i$ - $J_f$, where $J_i$ and $J_f$ are the zTAM of the initial and final electron states, respectively.

The decay width does not in general depend on the set of the wave-function of 
the radiated photon when the electron and photon eigen-states for a system are given.
For further study we calculate the decay width using the presently obtained photon wave-function.
A decay width of an electron can be calculated from the initial and final wave-functions of the electron in Eq.~(\ref{electron_function}) by the imaginary part of the electron self-energy.
The electron self-energy with an energy of $E$ is given by  
\begin{eqnarray}
\Sigma (\zr_1, \zr_2, E) 
 &=& 
i e^2 \int \frac{d p_0}{2 \pi}  
   \gamma_\mu S(\zr_1, \zr_2, p_0) \gamma_\nu  D^{\mu \nu} (\zr_1, \zr_2, E - p_0) ,
\end{eqnarray}
where $S$ and $D$ are the electron and photon propagators in the magnetic field:
\begin{eqnarray}
 S (\zr_1, \zr_2, p_0) &=& \sum_{L, n, h} \int \frac{d p_z}{ 2 \pi}
 \frac{ \psi (\zr_1 ; L, n, h, p_z) \psibar (\zr_2 ; L, n, h, p_z)}{ p_0 - E(L, n, h, p_z) + i \delta} ,
\\ 
D_{\mu \nu} (\zr_1, \zr_2, q_0) &=& \sum_{m, \alpha} \int \frac{d q_z d q_T q_T}{(2 \pi)^2} 
 \frac{A^{(\alpha)}_{m \mu} (\zr_{1}) A^{(\alpha) *}_{m \nu} (\zr_{2})}{q_0^2 -e_q^2 + i \delta} ,
\end{eqnarray}
where we omit the contribution from negative energy electrons in the electron propagator.
%
The decay width of the electron at an initial state $i$ is obtained as 
\begin{eqnarray}
\Gamma_e  (i) &=& 
- 2 \int d \zr_1 d \zr_2  
    \psibar_i(\zr_{1}) {\rm Im} \Sigma (\zr_1, \zr_2, E_i) \psi_i (\zr_{2})   
\nonumber \\ &=& 
\frac{e^2}{8 \pi^2} \sum_{f, m,\alpha}  \int \frac{d q_z d q_T q_T}{e_q} 
\frac{d p_{fz}}{ 2 \pi}
\delta (E_i - E_f - e_q )   
\left| \int d \zr \psibar_f (\zr) \Asla_m^{(\alpha) *} (\zr)  \psi_i(\zr) \right|^2 ,
\label{decay_width}
\end{eqnarray}
where $f$ indicates the final electron state. 
%
%
We rewrite the electron wave-function in Eq.~(\ref{electron_function}) and the photon field in Eq.~(\ref{photon_function}) as
\begin{equation}
\psi_b (\zr) = \frac{1}{\sqrt{R_z}} \phi_b (\zr_T) e^{i p_{bz} z}  \quad\mbox{and}\quad
A^{(\alpha)}_m (\zr) = V^{(\alpha)}_m (\zr_T) e^{i q_z z} ,
\label{rewirte}
\end{equation}
respectively.
Using Eqs.~(\ref{decay_width}) and (\ref{rewirte}),
the decay width  at a fixed final state  is written as 
\begin{eqnarray}
\Gamma_{if} &=&  
\frac{e^2}{8 \pi^2}  \int d q_z 
\left|  \int d \zr_T \phibar_f (\zr_T) \Vsla^{(\alpha) *} (\zr_T)  \phi_i (\zr_T) \right|^2 ,
\end{eqnarray}
where $E_f = \sqrt{ 2 eB N_{f} + (p_{iz} - q_z)^2 + m_e^2}$. 
For convenience, we define
\begin{eqnarray}
\cM( L_1, n_1 ; L_2, n_2 )  &=& 
\frac{1}{2} \int d^2 \zr  G^{L_1}_{n_1} (eB r^2/2)  J_{L_2 - L_1} \left( q_T r \right)
e^{i(L_1 - L_2)\phi} G^{L_2}_{n_2} (eB r^2/2) 
\label{MtEl}
\end{eqnarray}
and write 
$\cM_{22} = \cM(L_f, n_f ; L_i, n_i)$, $\cM_{11} = \cM(L_f-1, n_f ; L_i-1, n_i)$, $\cM_{21} = \cM(L_f, n_f ; L_i-1, n_i )$, and $\cM_{12} = \cM( L_f-1, n_f ; L_i, n_i)$.
We obtain
\bigskip
\begin{equation}
 \int d \zr_T \phibar_f (\zr_T) \Vsla^{(\alpha) *} (\zr_T)  \phi_i (\zr_T)
=
\sqrt{ \frac{(E_f+m_e) (E_i + m_e)}{4 E_i E_f } } \chi_{h_f}^\dagger \cT_\alpha \chi_{h_i},
\end{equation}
%
with
\begin{eqnarray}
\cT_{TM} &=& 
\frac{q_z}{e_q} \cM_{12} 
 \left[ \begin{array}{cc} \frac{p_{i T}}{E_i + m_e} 
                  & \frac{ i p_{i z} }{E_i + m_e}  - \frac{ i p_{f z}}{E_f + m_e} \\ 
0 &  \frac{ p_{f T} }{ E_i + m_e } 
\end{array} \right] 
- \frac{q_z}{e_q} \cM_{21} 
 \left[ \begin{array}{cc} - \frac{p_{f T}}{E_f + m_e} & 0\\ 
\frac{ i p_{f z} }{E_f + m_e}  - \frac{ i p_{i z}}{E_i + m_e }  & - \frac{ p_{iT} }{ E_i + m_e } 
\end{array} \right] 
%
\nonumber \\ &&  
+ \frac{2q_T}{e_q}\cM_{11} \left[ \begin{array}{cc}  
 \frac{p_{iz}}{E_i + m_e} + \frac{p_{fz}}{E_f + m_e}  & - \frac{ i p_{i T} }{ E_i + m_e }  \\ 
  \frac{i p_{fT} }{E_f + m_e} & 0  
\end{array} \right]
+ \frac{2 q_T}{e_q} \cM_{22} \left[ \begin{array}{cc}  0  &  \frac{i p_{fT}}{E_f + m_e}  \\ 
-  \frac{ i p_{iT}}{E_i + m_e} &  \frac{p_{iz}}{E_i + m_e} + \frac{p_{fz}}{E_f + m_e}  
\label{cT1}
 \end{array} \right]  ,
 %
\\ 
\cT_{TE} &=& 
 \cM_{12}
 \left[ \begin{array}{cc} \frac{p_{i T}}{E_i + m_e} 
                  & \frac{ i p_{i z} }{E_i + m_e}  - \frac{ i p_{f z}}{E_f + m_e} \\ 
0 &  \frac{ p_{f T} }{ E_i + m_e } 
\end{array} \right] 
+  \cM_{21}
 \left[ \begin{array}{cc} - \frac{p_{f T}}{E_f + m_e} & 0 \\ 
\frac{ i p_{f z} }{E_f + m_e}  - \frac{ i p_{i z}}{E_i + m_e}  & - \frac{ p_{iT} }{ E_i + m_e } 
\end{array} \right] ,
\label{cT2}
\end{eqnarray}
where $p_{i(f)T} = \sqrt{2eB  N_{i(f)}}$. 
%
Because the electron spin is not a good quantum number in the relativistic framework,
we make the average of the strength in Eq.~(\ref{decay_width}) for the initial spin and 
the summation for the final spin.
We finally obtain the decay width of
\begin{eqnarray}
\frac{d \Gamma^{(a)}_{if}}{d p_{fz} } = \frac{d \Gamma_{if}^{(a)} }{d q_{z}  }
&=&
\frac{e^2}{8 \pi^2} 
\left|  \int d \zr_T \phibar_f (\zr_T) \Vsla^{(\alpha) *}_m (\zr_T)  \phi_i (\zr_T) \right|^2 .
\nonumber \\
&=& \frac{\alpha_e}{2 \pi} \frac{(E_f + m_e) (E_i + m_e) }{ E_i E_f } {\rm Tr} \left[ \cT_a \cT_a \right] ,
\label{integrated_decay_width} 
\end{eqnarray}
The possible momentum of an emitted photon with $m$ is limited by 
\begin{equation}
\frac{eBm}{ E_i + |p_{iz}| } \le e_q \le \frac{eBm}{ E_i - |p_{iz}| } .
\label{energy_region}
\end{equation}
The maximum (minimum) energy for a radiation with $m$ is proportional to $m$, 
and the average energy at $m$ increases as $m$ increases.


To investigate radiations from magnetized neutron stars we numerically calculate the decay widths of electrons under strong magnetic fields of 10$^{12}$~G and 10$^{13}$~G.
Figure~\ref{TWid} shows the decay widths of electrons as a function of $m$ using Eq.~(\ref{integrated_decay_width}).
The results are multiplied with the gamma factor along the $z$-direction, 
$\gamma_z = E_i / \sqrt{E_i^2 - p_{iz}^2}$, 
so that the results are independent of the z-component of the initial electron momentum of $p_{iz}$ when $|p_{iz}| \gg m_e$.
To obtain the total energy spectrum and the fraction of photon vortices of radiated photons from an electron in an initial state, we calculate the decay widths to individual final states for various $m$ values.
In Fig.~\ref{WdQz}(a) we present the total energy spectrum and energy spectra of individual modes of photon vortices with $m$ in synchrotron radiations from electrons with an energy of 50~MeV, which is integrated over all the radiation angle.

\section{Discussion and conclusion}

We have finally obtained the wave-function of the radiated photon as Eq.~(\ref{photon_function}).
The $m$-th harmonic radiation is the eigen-state of the zTAM $m$.  
This is consistent with the previous results \cite{Katoh17, Bogdanov18, Bogdanov19}.
Though in Ref.~\cite{Bogdanov18, Bogdanov19}  the photon wave-functions were chosen to be  the eigen-states of the helicity in the limit of the photon transverse momentum over the photon energy $q_T / e_q =0$, we have chosen them to be the eigen-state of TM and TE [see Eq.~(\ref{photon_function})] \cite{Jaurgui05}.
To examine the choice of the basis we show the contributions from TM and TE states   
and those from the helicity $s=+1$ and $s= -1$ states as the function of $q_T/e_q$ in  Fig.~\ref{fig6}.
When $q_T \lesssim 0.01 e_q$ the state with $s=+1$ dominantly contributes to the decay width. However, as $q_T$ increases  the contribution from $s=-1$ becomes larger.
In contrast, the TM state in Eq.~(\ref{photon_function}) dominates when $q_T \gtrsim 0.02 e_q$.
Although the basis of the helicity is useful for calculating in the condition of $q_T \ll e_q$, the wave-function as the eigen-state of TM and TE can give more precise results for wide region of $q_T$ under extremely strong magnetic fields.
Furthermore, the previous studies \cite{Sasaki08,Bogdanov18, Bogdanov19} calculated the synchrotron radiation from helical undulators, which consists of different direction magnets. If it is calculated using Landau quantization, its result is expected be different from the present result for uniform magnetic fields.

The fraction of photon vortices with $m$ $\ge$ 2 in the total flux depends on the magnetic field strength and the initial angular momentum.
As the magnetic strength becomes stronger, the fraction of photon vortices increases. Furthermore, as shown in Fig.~\ref{TWid} the fraction of photon vortices increases with increasing the initial electron angular momentum, which is proportional to the square of the diameter of the electron spiral motion.
The highly linear polarization of $\gamma$-rays from GRBs indicate a magnetized baryonic jet with large-scale uniform strong magnetic fields \cite{Mundell13}.
However, the present calculation does not depend on the large-scale structure of a magnetic field.
It is considered that magnetic fields in astrophysical environments are locally homogeneous for photon vortex generation compered with electron spiral motion diameters. 
The diameter of a spiral motion of an electron is estimated to be
$d \sim 1.6 \times \sqrt{ 10^{13}/B~[\mbox{G}]  }L_{i}^{1/2}~[\mbox{pm}].$
In the present assumed conditions, the diameters are shorter than 10$^{-8}$~cm.
Thus, the uniform magnetic field is a good approximation for calculation of synchrotron radiation in astrophysical objects.

The possible energies of the fundamental radiation and $m$-th harmonic radiations are restricted by the region expressed by Eq.~(\ref{energy_region}).
As shown in Fig.~\ref{WdQz}(a) the upper limit of the energy of the fundamental radiations is 4.1~MeV when the initial electron energy is approximately 50~MeV.
Therefore, below 4.1 MeV both the fundamental radiation and the high-order harmonic radiations are emitted, whereas in the energy region of 4.1$\--$50~MeV only photon vortices are radiated.
This trend is seen in spectra in the wide energy region as shown in Figs.~\ref{WdQz}(b) and ~\ref{WdQz2}(a,~b).
In astrophysical environments, the energy distribution of thermal electrons follows Fermi-Dirac distribution.
High energy $\gamma$-rays observed in afterglow phases of GRBs suggest non-thermal electrons whose energy distribution is described by power law \cite{Granot02}. 
In both cases the spectral density of electrons decreases with increasing electron energy in high energy region.
With the present result that the only photon vortices are generated in the energy region higher than the upper limit of the fundamental radiation, it is expected that the fraction of photon vortices increases with increasing energy and photon vortices are predominantly produced in the high energy region.

The observation of polarized X/$\gamma$-rays with detectors onboard satellites and interplanetary space explores \cite{Kalemci07, Yonetoku12, Mundell13, Wiersema14} suggests that photon vortices could be also observed in the space near the earth.
A method to measure Laguerre-Gaussian light at optical wavelengths from astrophysical objects has been proposed \cite{Berkhout08}.
The present calculation predicts photon vortex generation but does not predict any coherency. 
This means that the observed light seems to be white light and photon vortices should be identified using a method based upon a quantum phenomenon.
At present, linearly polarized $\gamma$/X-rays are measured using detectors based upon Compton scattering.
To measure $\gamma$-rays with a wave-function of Laguerre-Gaussian \cite{Maruyama19a} or Hermite-Gaussian \cite{Maruyama19b} we have proposed the use of Compton scattering.
Because the presently obtained wave-function in Eq.~(\ref{photon_function}) has the feature similar to that of the Laguerre-Gaussian photon \cite{Maruyama19a}, the measurements of photon vortices generated by the synchrotron radiations are probably possible in similar manners.

Photon vortices may affect stellar nucleosyntheses. 
It is known that neutron-deficient isotopes of heavy elements are synthesized by photodisintegration reactions in supernova explosions ($\gamma$-process) \cite{Woosley78, Hayakawa04}.
Giant dipole resonance (GDR) is the dominant reaction between photons and nuclei in the energy region of 10$\--$30 MeV.
However, it was pointed out that, if a photon vortex with large total angular momentum of $J$ $\ge$ 2 incidents on an even-even nucleus of a spin and parity of $J^{\pi}$ = 0$^+$, the excitation to states with $J^{\pi}$ = 1$^{-}$ through GDR is forbidden because of the conservation law of angular momentum \cite{Taira17}.
Thus, when nucleosyntheses associated with photodisintegration reactions occurs in a strong magnetic field, the isotopic abundance distribution of a synthesized element is expected to be different form the solar abundances.
The r-process paths in the magnetohydrodynamical supernovae \cite{Nishimura06} and in jets in collapsars \cite{Nakamura15} shifts toward more neutron rich regions by photon vortices.
The isotopic abundances affected by photon vortices may be observed in presolar grains \cite{Lodders05}, which record individual nucleosyntheses before the solar system formation.

\section{Summary}

The difference between calculations with/without Landau quantization for strong magnetic fields becomes large increasing magnetic field strengths.
In the present study we have calculated the wave-function of photon vortices radiated from electrons under Landau levels.
The wave-function is the eigen-state of the $z$-component of the total angular momentum when the electron has a spiral motion along $z$-axis.
This is consistent with the previous results.
The photon vortices, in principle, predominantly produced in astrophysical objects with extremely strong magnetic fields as high as 10$^{12}{\--}$10$^{13}$~G.
These photons have not any coherent structure so that they seem to be like white light. 
However, they are expected to be measured by a new detector based upon Compton scattering onboard satellites in future.
Photonuclear reactions with photon vortices are hindered because of the conservation law of angular momentum so that photon vortices change the isotopic abundances of synthesized nuclide in astrophysical environments with strong magnetic fields.

\bigskip
\noindent{\bf Acknowledgements}

This work was supported by Grants-in-Aid for Scientific Research of 
JSPS (JP20K03958, JP19K03833, JP18H03715) and 
the grant of Joint Research by the National Institutes of Natural Sciences (NINS), 
(NINS program No, 01111701).
MKC work was supported by the National Research Foundation of Korea (Grant
Nos. NRF-2020R1A2C3006177 and NRF-2013M7A1A1075764).




\begin{thebibliography}{00}





\bibitem{MacFadyen99}
A. I. MacFadyen, S. E.Woosley, Astrophys. J. {\bf 524},  (1999) 262$\--$289.
https://iopscience.iop.org/article/10.1086/307790
\bibitem{Ardeljan00}
 N. V. Ardeljan, G. S. Bisnovatyi-Kogan, S. G.  Moiseenko, Astron. Astrophys. {\bf 355},  (2000) 1181$\--$1190.
http://adsabs.harvard.edu/full/2000A\%26A...355.1181A

\bibitem{Takiwaki11}
T. Takiwaki, K. Kotake,  Astrophys. J. {\bf 743}, (2011) 30.
https://iopscience.iop.org/article/10.1088/0004-637X/743/1/30

\bibitem{Mereghetti08}
S. Mereghetti, Annu. Rev. Astrophysm. {\bf 15}, (2008) 225.
https://link.springer.com/article/10.1007

\bibitem{Zhang10}
D. Zhang, Astrophys. J. {\bf 718},  (2010) 841$\--$846.
https://iopscience.iop.org/article/10.1088/0004-637X/718/2/841

\bibitem{McKinney13}
J. C. McKinney, A. Tchekhovskoy, R. D. Blandford, Science, {\bf 39},  (2013) 49$\--$52.
https://science.sciencemag.org/content/339/6115/49?sid=130a9c47-0571-4a47-8be3-bd58afd58afc

\bibitem{Lei18}
W. Lei, B.  Zhang, X. Wu, E. Liang, Astrophys. J.  {\bf 849}, (2017) 47.
https://iopscience.iop.org/article/10.3847/1538-4357/aa9074

\bibitem{Coburn03}
W. Cobrun, S. E. Boggs, Nature, {\bf 423},  (2003) 415$\--$417.
https://www.nature.com/articles/nature01612

\bibitem{Wigger04}
C. Wigger, W. Hajdas, K. Arzner, M. G{\"u}del, A. \& Zehnder, Astrophys. J. {\bf 613}, (2004) 1088$\--1100$.
https://iopscience.iop.org/article/10.1086/423163/meta

\bibitem{Kalemci07}
E. Kalemci, S. E. Boggs, C. Kouveliotou, M. Finger and M. G. Baring, Astrophys. J. Supp. {\bf 169} (2007) 75$\--$82,
https://iopscience.iop.org/article/10.1086/510676

\bibitem{Yonetoku12}
D. Yonetoku, et al., Astrophys. J. Lett. {\bf 758} (2012) L1, 
https://iopscience.iop.org/article/10.1088/2041-8205/758/1/L1

\bibitem{Panaitescu00}
A. Panaitescu, P. M{\'e}sz{\'a}ros, Astrophys. J.  544 (2000) L17$\--$L21.
https://iopscience.iop.org/article/10.1086/317301

\bibitem{Chang14}
Z. Chang, H. Lin, Y. Jiang, Astrophys. J. {\bf 783} (2014) 30, 
https://iopscience.iop.org/article/10.1088/0004-637X/783/1/30/meta

\bibitem{Mao18}
J. Mao, S. Covino, J. Wang, Astrophys. J.  {\bf 860}, (2018) 153.
https://iopscience.iop.org/article/10.3847/1538-4357/aac5d9


\bibitem{Sokolov45}
A. A. Sokolov, J. Phys. USSR, {\bf 9} (1945) 363.

\bibitem{Sokolov53}
A. A. Sokolov and I. M. Ternov, Zh. Eksp. Teor. Fiz. {\bf 25} (1953) 698.

\bibitem{Klepikov54}
N. Klepikov,  Zhur. Eksptl.’i Teoret. Fiz. {\bf 26} (1954) 59.

\bibitem{Bezchastnov88}
V. G. Bezchastnov, G. Pavlov, Astrophys. J. Supp. {\bf 148},  (1988) 257.
https://link.springer.com/article/10.1007/BF00645965

\bibitem{Pavlov91}
G. G. Pavlov, V. G. Bezchastnov, P.  Mészáros, S. G. Alexander,  Astrophys. J. {\bf 380}, (1991) 541$\--$549.
http://adsabs.harvard.edu/full/1991ApJ...380..541P



\bibitem{Bagrov}
V. G. Bagrov, Quantum Theory of Synchrotron Radiation, in: V. A Bordovitsyn (Ed.) Synchrotron Radiation Theory and Its Development, World Scientific Publishing Co Pte Ltd., 1999, pp. 121$\--$178. 

\bibitem{Dorofeyev}
O. F. Dorofeyev, A. V. Borisov, and V. Ch. Zhukovsky, Synchrotron Radiation in a Strong Magnetic Field, in V. A Bordovitsyn (Ed.) Synchrotron Radiation Theory and Its Development, World Scientific Publishing Co Pte Ltd., 1999, pp. 347$\--$383.

\bibitem{Bisnovatyi-Kogan}
G. S. Bisnovatyi-Kogan, Synchrotron Radiation in Astrophysics, in: V. A Bordovitsyn (Ed.) Synchrotron Radiation Theory and Its Development, World Scientific Publishing Co Pte Ltd., 1999, pp. 385$\--$443.

\bibitem{Wistisen15}
T. N. Wistisen, Phys. Rev. D {\bf 92}, (2015) 045045.
https://journals.aps.org/prd/abstract/10.1103/PhysRevD.92.045045

\bibitem{Kruining19}
K. V. Kruining, F. Mackenroth, J. B. Götte, Phys. Rev. D {\bf 100},  (2019) 056014, 
https://journals.aps.org/prd/abstract/10.1103/PhysRevD.100.056014




\bibitem{Maruyama16} 
T. Maruyama, M.~-K. Cheoun, T. Kajino, and G.~J. Mathews, Phys. Lett. B {\bf 75},  (2016) 125.
https://doi.org/10.1016/j.physletb.2016.03.065

\bibitem{Allen92} 
L. Allen, M.~W. Beijersbergen, R.~J.~C. Spreeuw and J.~P. Woerdman, Phys. Rev. A {\bf 45} (1992) 8185-8189, 
https://journals.aps.org/pra/abstract/10.1103/PhysRevA.45.8185

\bibitem{Harwit03}
M. Harmit, Astrophys. J. {\bf 597} (2003) 1266$\--$1270, 
https://iopscience.iop.org/article/10.1086/378623

\bibitem{Elias08}
N.~M. Elias II, Astronom. Astrophys. {\bf 492} (2008) 883$\--$992,	
 	https://doi.org/10.1051/0004-6361:200809791 
 	
\bibitem{Berkhout08}
G.~C.~G. Berkhout, M.~W. Beijersbergen, Phys. Rev. Lett. {\bf 101}  (2008) 100801, 
https://journals.aps.org/prl/abstract/10.1103/PhysRevLett.101.100801

\bibitem{TTMA11}
F. Tamburini, Bo Thide, Nature Phys. {\bf 7} (2011) 195$\--$197,
https://www.nature.com/articles/nphys1907

\bibitem{Jentschura11a}
U.~D. Jentschura, V.~G. Serbo, Phys. Rev. Lett. {\bf 106}  (2011) 013001, 
https://journals.aps.org/prl/abstract/10.1103/PhysRevLett.106.013001

\bibitem{Harris15}
J. Harris, V. Grillo, E. Mafakheri, G. C. Gazzadi, S. Frabboni, R. W. Boyd, and E. Karimi, Nature Phys. {\bf 11} (2015) 629$\--$634.
https://www.nature.com/articles/nphys3404

\bibitem{Petrillo16}
V. Petrillo, G. Dattoli, I. Drebot, and F. Nguyen, Phys. Rev. Lett. {\bf 117},  (2016) 123903.
https://pubmed.ncbi.nlm.nih.gov/27689277/

\bibitem{Sherwin17}
J.~A. Sherwin, Phys. Rev. A {\bf 95},  (2017) 052101.
https://doi.org/10.1103/PhysRevA.95.052101

\bibitem{Afanasev17}
A. Afanasev, V.~G. Serbo, M. J. Phys. G Nucl. Part. Phys. {\bf 45}  (2018) 055102, 
http://iopscience.iop.org/article/10.1088/1361-6471/aab5c5/

\bibitem{Taira17}
Y. Taira, T. Hayakawa, M. Katoh, Sci. Rep. {\bf 7} (2017) 5018, 
https://doi.org/10.1038/s41598-017-05187-2


\bibitem{Peshkov18}
A.~A. Peshkov, A.~ V. Volotka, A. Surzhykov, and S. Fritzsche, Phys. Rev. A {\bf 97},  (2018) 023802,
https://doi.org/10.1103/PhysRevA.97.023802

\bibitem{Maruyama19a}
T. Maruyama, T.  Hayakawa, and T. Kajino, Sci. Rep. {\bf 9},  (2019) 51.
https://www.nature.com/articles/s41598-018-37096-3

\bibitem{Lee19}
J. C. T. Lee, S. J. Alexander, S. D.  Kevan, S.  Roy, B. J.  McMorran, Nature Photonics, {\bf 13},  (2019) 205$\--$209.
https://www.nature.com/articles/s41566-018-0328-8

\bibitem{Sherwin20}
J. A. Sherwin, Phys. Rev. Res. {\bf 2},  (2020) 013168.
https://doi.org/10.1103/PhysRevResearch.2.013168

\bibitem{Sasaki08}
S. Sasaki,  I. McNulty,  Phys. Rev. Lett. {\bf 100}  (2008) 124801, 
https://doi.org/10.1103/PhysRevLett.100.124801

\bibitem{Bahrdt13}
J. Bahrdt, et al., Phys. Rev. Lett.  {\bf 111} (2013)  034801, 
https://doi.org/10.1103/PhysRevLett.111.034801

\bibitem{Hemsing13}
E. Hemsing, A. Knyazik, M. Dunning, D. Xiang, A. Marinelli, C. Hast, and J.~B. Rosenzweig, Nature Phys. {\bf 9},  (2013) 549.
https://www.nature.com/articles/nphys2712

\bibitem{Katoh17}
M. Katoh, et al., Phys. Rev. Lett. {\bf 118}  (2017) 094801, 
https://doi.org/10.1103/PhysRevLett.118.094801

\bibitem{Bogdanov18}
O. V. Bogdanov, P. O. Kazinski, G. Yu. Lazarenko, Phys. Rev. A. {\bf 97},   (2018) 033837.
https://doi.org/10.1103/PhysRevA.97.033837

\bibitem{Bogdanov19}
O. V. Bogdanov, P. O. Kazinski, G. Yu. Lazarenko, Phys. Rev. D. {\bf 99},  (2019) 116016.
https://doi.org/10.1016/j.aop.2019.03.028

\bibitem{RKubo65}
R. Kubo, S.~J. Miyake, N. Hashitsume,  Solid State Phys. {\bf 17},  (1965) 269.
https://doi.org/10.1016/S0081-1947(08)60413-0

\bibitem{Jaurgui05}
R. J{\'a}urgui and S. Hacyan, Phys. Rev. A. {\bf 71}, (2005) 033411.
https://doi.org/10.1103/PhysRevA.71.033411

\bibitem{Mundell13}
C.~G. Mundell, et al., Nature, {\bf 504},  (2013) 119.
https://www.nature.com/articles/nature12814

\bibitem{Wiersema14}
K. Wiresema, et al., Nature. {\bf 509},  (2014) 201
https://www.nature.com/articles/nature13237

\bibitem{Granot02}
J. Granot, R. Sari, Astrophys. J. {\bf 568},  (2002) 820$\--$829.
https://iopscience.iop.org/article/10.1086/338966

\bibitem{Maruyama19b}
T. Maruyama, T.  Hayakawa,  T. Kajino, Sci. Rep. {\bf 9},  (2019) 7998.
https://www.nature.com/articles/s41598-019-44120-7


\bibitem{Woosley78}
S. E. Woosley, W. M. Howard, Astrophys. J. Supp. {\bf 36}, (1978) 285$\--$304 .
https://ui.adsabs.harvard.edu/abs/1978ApJS...36..285W/abstract

\bibitem{Hayakawa04}
T. Hayakawa, et al., {\it Phys. Rev. Lett.} {\bf 93},  (2004) 161102.
https://doi.org/10.1103/PhysRevLett.93.161102

\bibitem{Nishimura06}
S. Nishimura, et al.  {\it Astrophys. J.} {\bf  642},  (2006) 410$\-–$419.
https://iopscience.iop.org/article/10.1086/500786

\bibitem{Nakamura15}
K. Nakamura, T. Kajino, G. J. Mathews, S. Sato, S. Harikae,  {\it Astron. Astrophys.} {\bf 582},  (2015) A34.
https://doi.org/10.1051/0004-6361/201526110 

\bibitem{Lodders05}
K. Lodders, S. Amari, {\it Geochemistry}, {\bf 65},  (2005) 93$\--$166.
https://doi.org/10.1016/j.chemer.2005.01.001

\end{thebibliography}


\newpage

\begin{figure}[htb]
\begin{center}
\vspace{1em}
\includegraphics[scale=0.45,angle=0]{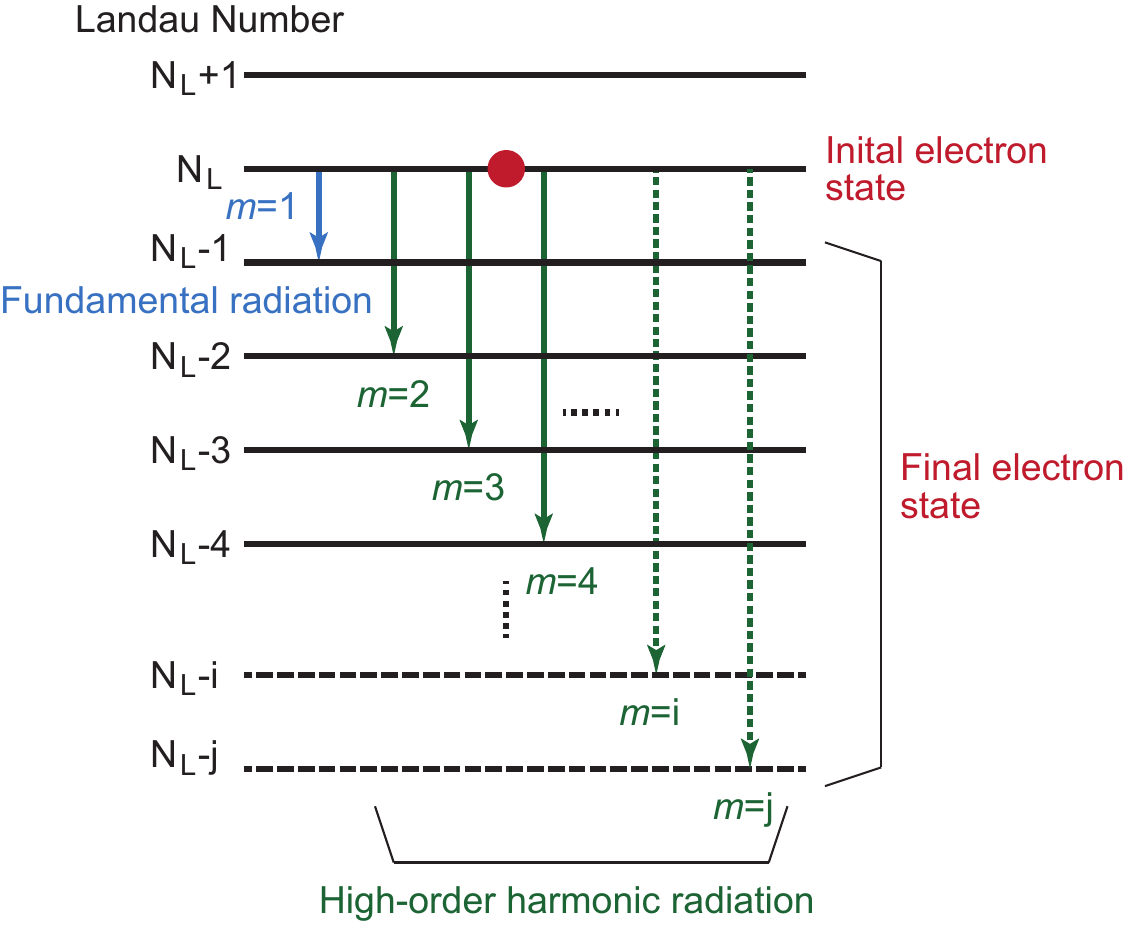}
\caption{\small
Schematic view for electron state in a Landau level and decay modes.
}
\label{Landau_level}
\end{center}
\end{figure}

\begin{figure}[htb]
\begin{center}
\vspace{1em}
{\includegraphics[scale=0.65,angle=270]{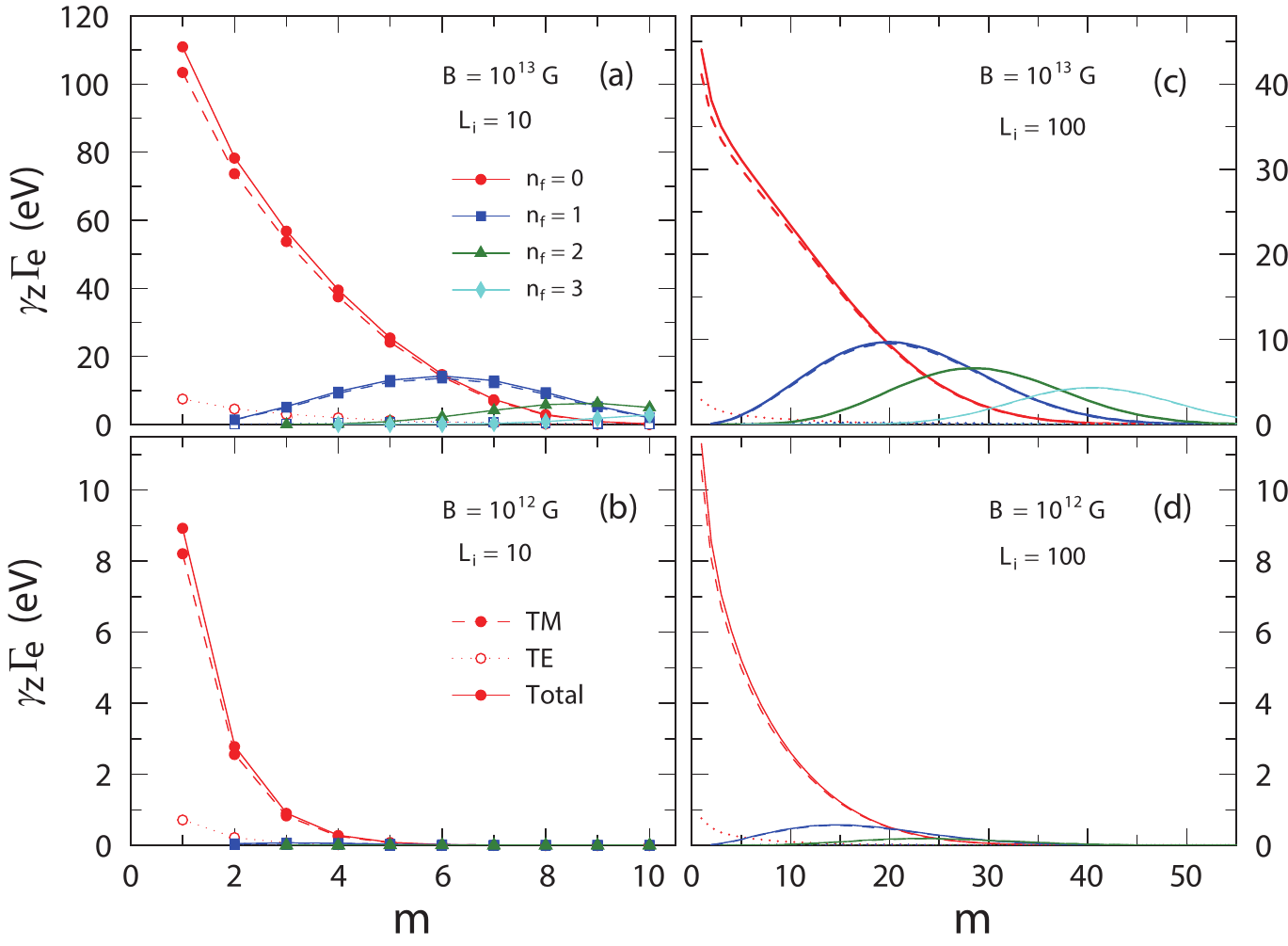}}
\caption{\small
Decay widths of electrons with $n_i=0$ as a function of zTAM of an emitted photon when the final electron Landau number is fixed.
The initial orbital angular momenta are $L_i = 10$ ($J_i = 10-1/2$) for (a,b) and $L_i = 100$ ($J_i = 100-1/2$) for (c,d).
The magnetic field strengths are $B=10^{13}$G for (a,c) and $B=10^{12}$G for (b,d).
The dashed and dotted lines represent the results of the photon field at the state-1 and the state-2, respectively, and the solid line indicates a summation of the two contributions.
}
\label{TWid}
\end{center}
\end{figure}

\begin{figure}[htb]
\begin{center}
\vspace{1em}
{\includegraphics[scale=0.55,angle=0]{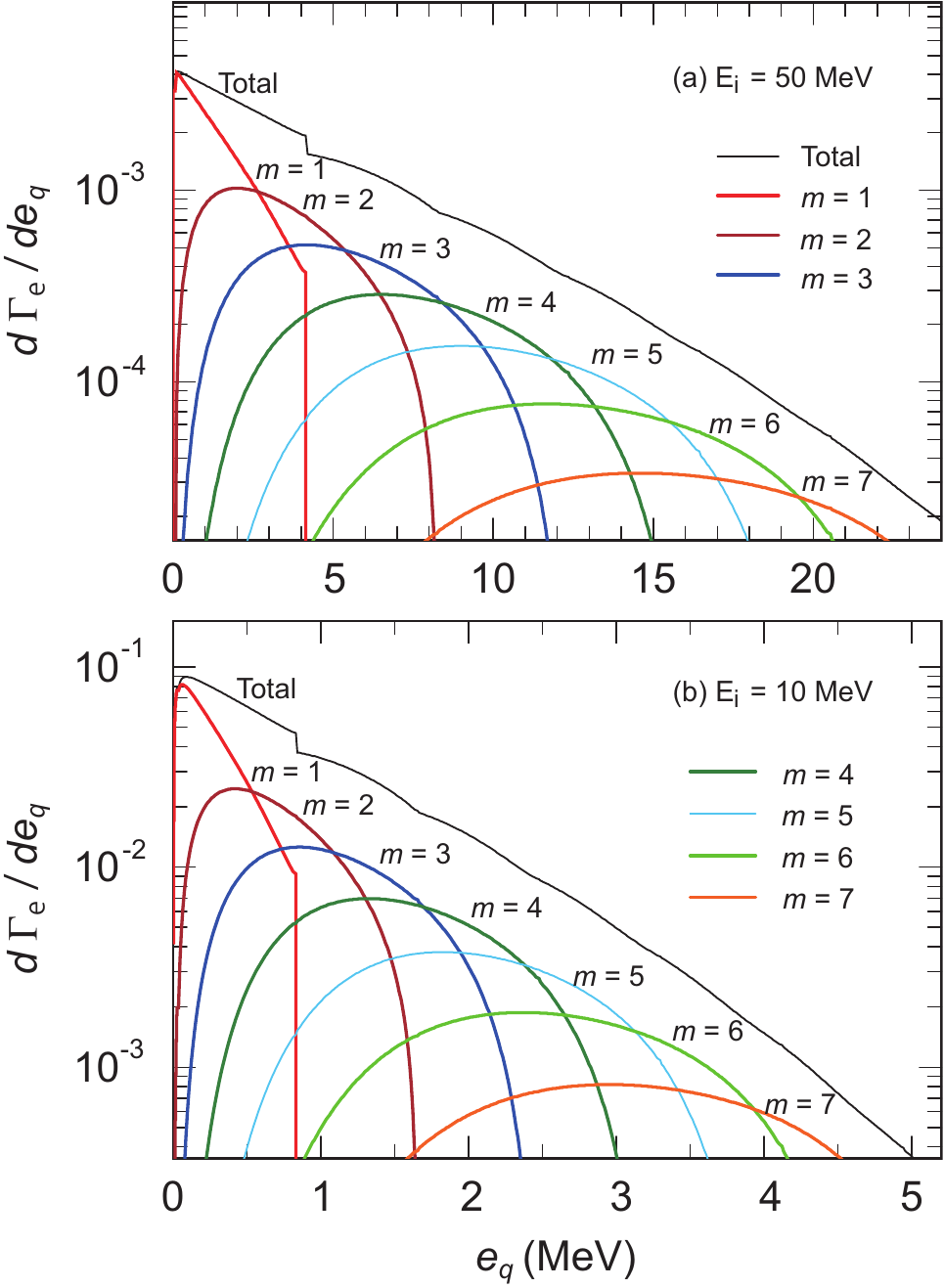}}
\caption{\small
$d \Gamma_e / d e_q$ with $n_i = 0$, $L_i = 10$, and $p_{iz} = 50$~MeV (a) or 10~MeV (b) at the magnetic field strength $B = 10^{13}$G
as a function of the $z$-component of the photon momentum, $q_z$.
}
\label{WdQz}
\end{center}
\end{figure}

\begin{figure}[htb]
\begin{center}
\vspace{1em}
{\includegraphics[scale=0.55,angle=0]{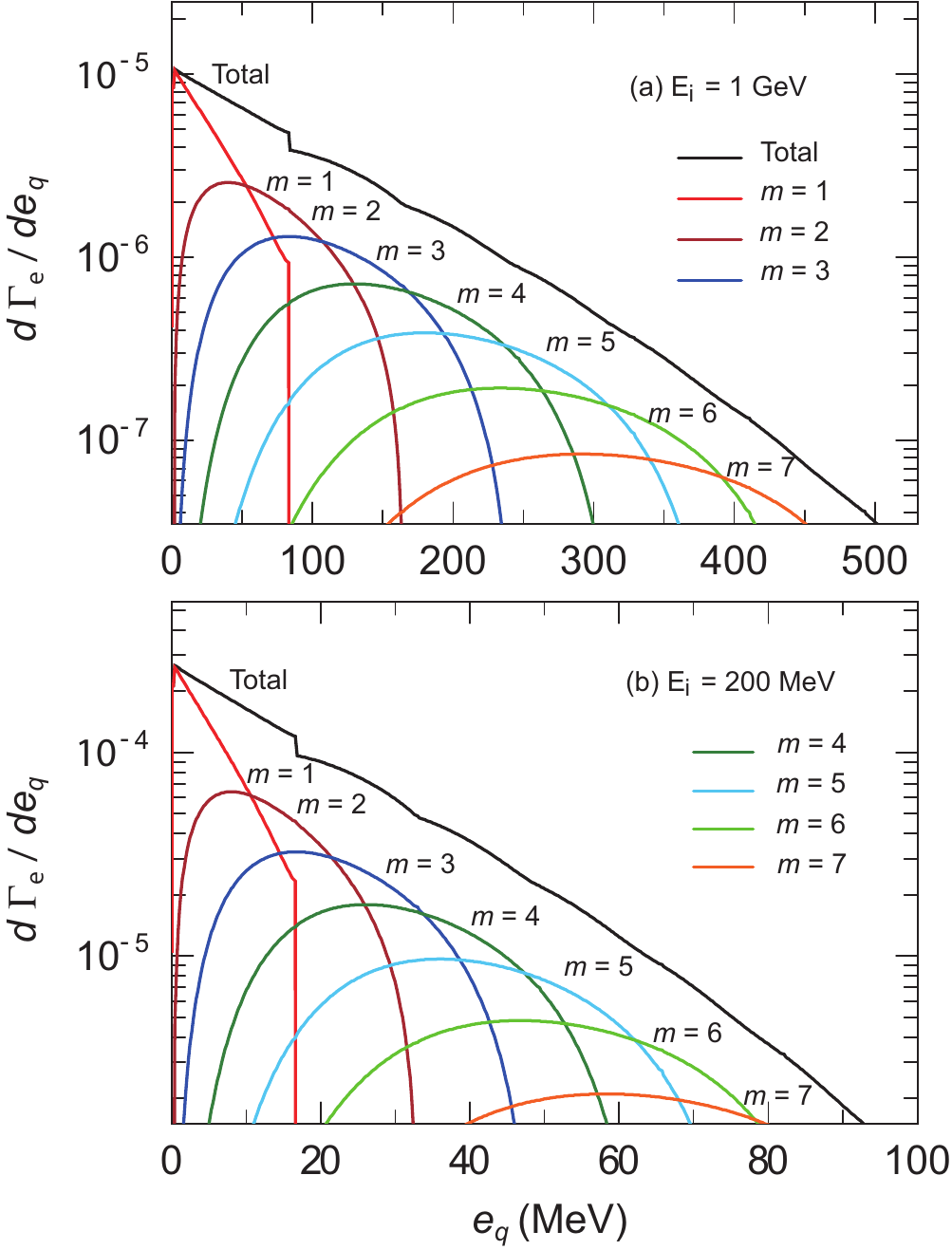}}
\caption{\small
$d \Gamma_e / d e_q$ with $n_i = 0$, $L_i = 10$, and $p_{iz} = 1$~GeV (a) or 200~MeV (b) at the magnetic field strength $B = 10^{13}$G
as a function of the $z$-component of the photon momentum, $q_z$.
}
\label{WdQz2}
\end{center}
\end{figure}

\begin{figure}[htb]
\begin{center}
\vspace{1em}
{\includegraphics[scale=0.5,angle=0]{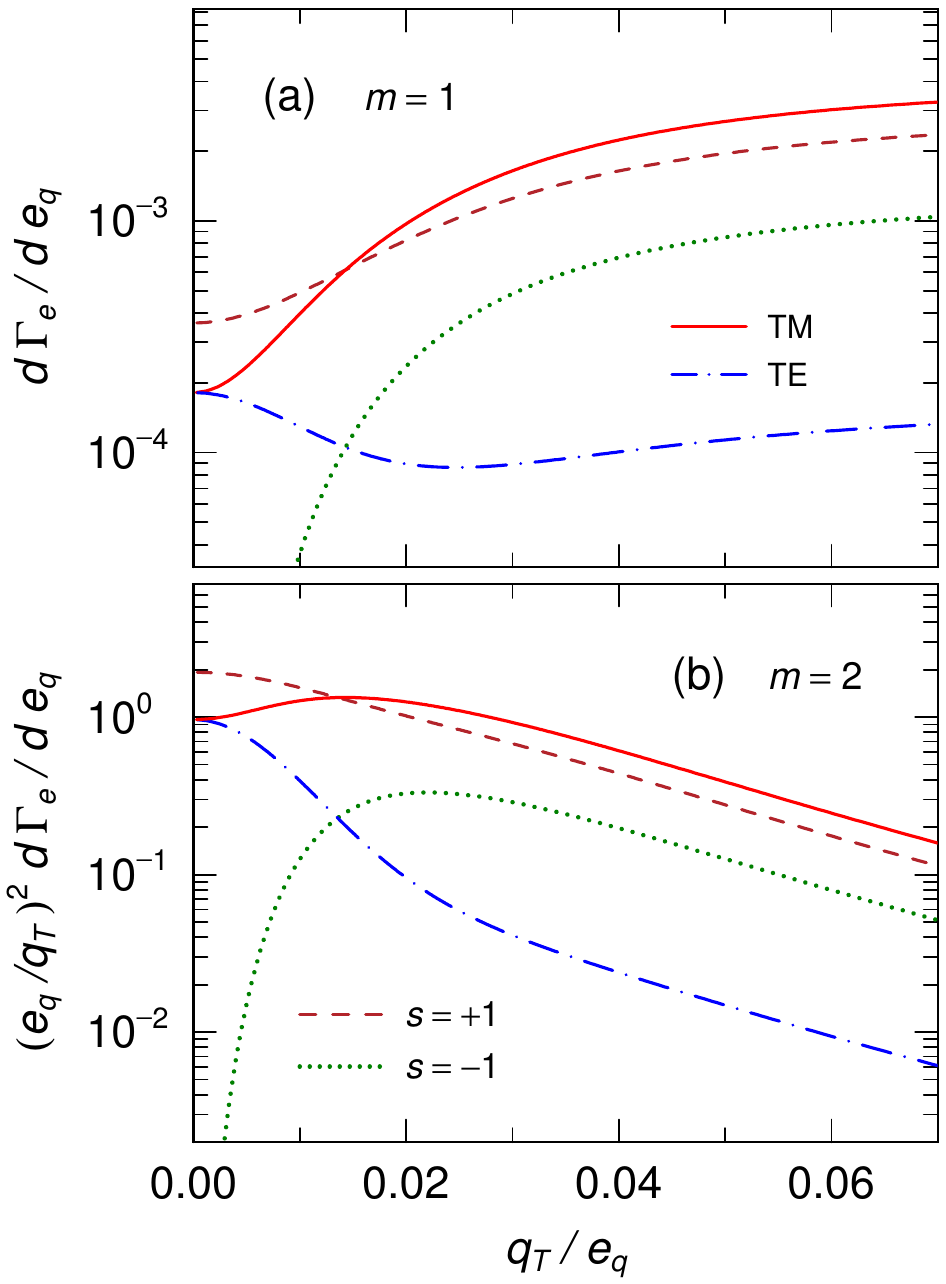}}
\caption{\small
$d \Gamma_e / d e_q$ with $n_i = 0$, $L_i = 10$, $p_{iz} = 50$~MeV, and $m$ = 1 (a) or $m$ = 2 (b) at the magnetic field strength $B = 10^{13}$G as a function of $q_T$/$e_q$.
The solid , dot-dashed, dashed and dotted lines represent the contributions from TM, TE, $s=+1$ and $s=-1$ states for final phorons.
}
\label{fig6}
\end{center}
\end{figure}

\end{document}